# From amorphous speckle pattern to reconfigurable Bessel beam via wavefront shaping

Diego Di Battista,[1,*] Daniele Ancora,[1] Marco Leonetti,[2] and Giannis Zacharakis[1]

[1]Institute of Electronic Structure and Laser, Foundation for Research and Technology-Hellas, N. Plastira 100, VasilikaVouton, 70013, Heraklion, Crete, Greece
[2] Center forLife Nano Science@Sapienza, Instituto Italiano di Tecnologia, Viale Regina Elena, 291 00161 Rome (RM), Italy
*Corresponding author: dibattista.d@iesl.forth.gr

**Bessel beams are non-diffracting light structures, which can be produced with simple tabletop optical elements such as axicon lenses or ring spatial filters and coherent laser beams. One of their main characteristic is that Bessel beams maintain their spatial characteristics after meters of propagation. In this paper we demonstrate a system and method for generating Bessel beams from amorphous speckle patterns, exploiting adaptive optimization by a spatial light modulator. These speckles are generated by light modes transmitted through a scattering curtain and selected by a ring shaped filter. With the proposed strategy it is possible to produce at user defined positions, reconfigurable, non-diffracting Bessel beams through a disordered medium.**

During the last decade biomedical imaging has been revolutionized by the advances of optical technologies. The adaptation of various innovative methodologies is ever increasing and has allowed to overcome fundamental limits in optical imaging. The plethora of new systems and methods has changed the standards in biological imaging from the traditional microscope to more hollistic approaches covering the wide range from molecules to cells to whole organisms. Technologies such as STED, PALM and STORM have pushed the limits down to the nanoscale resolution [1,2]. On the other hand, the adaptation of optoacoustics has for the first time allowed to image in high resolution through turbid tissues in depth of up to a few centimeters. Furthermore, adaptive optics and wavefront shaping have been used to compensate for light diffusion through highly disordered media and are becoming increasingly relevant to biomedical, and biological experiments ranging from imaging to therapy [3]. Light manipulation through scattering media has been implemented in numerous other fields including optical focusing [4], imaging correction [5], light polarization [6], sub-diffractive microscopy [7], non-linear system tuning [8] and communication control [9].

Indeed, light may be focused through a heavily scattering curtain by correcting the wavefront with a Spatial Light Modulator (SLM): a scattering material plus an adaptive optical system is usually named an opaque lens. Opaque lenses can be used for sub-diffractive resolution through a scattering layer [7] and to generate light foci at user defined positions [4,10].

Furthermore, wavefront correction has been exploited for enhancing the quality of Bessel beams [11,12] for non-diffractive illumination [13,14]. Such approaches can benefit fluoroscence imaging techniques due to the increased depth-of-field with a drastic improvement in image quality, increased acquisition framerate and reduction of acquisition time [15].

Bessel beams have the property to be "self-reconstructing" meaning that the light-structure is immune to scattering encountered along the beam trajectory [16], (thus enabling applications such as deep laser writing [17]), but not to extended scattering barriers [18].

In this Letter we describe a method for producing Axicon Opaque Lenses (AOL) able to combine the properties of non-diffractive Bessel beams with the advantages of adaptive focusingenabling large depth of field combined with user defined reconfigurability.

By filtering the spatial spectrum of the light transmitted at the back of a scattering sample [19] with a ring-shape spatial filter we create an amorphous speckle pattern [20,21,22]. Amorphous speckle are non-random distributions charcterized by a short-range order spatial resolution. By correcting in phase the wavefront of the incident beam it is possible to manipulate the interference pattern generated by the superposition between the fields selected by the ring-shape filter. Using the intensity on a defined region of the camera as feedback we can focus on a specific target.

The experimental setup used in our studies is shown in schematic representation in Figure 1. A He-Ne laser emitting at 594nm is used as a source, while the beam is magnified by a10x telescope and corrected only in phase by a SLM (Pluto, Holoeye Germany).

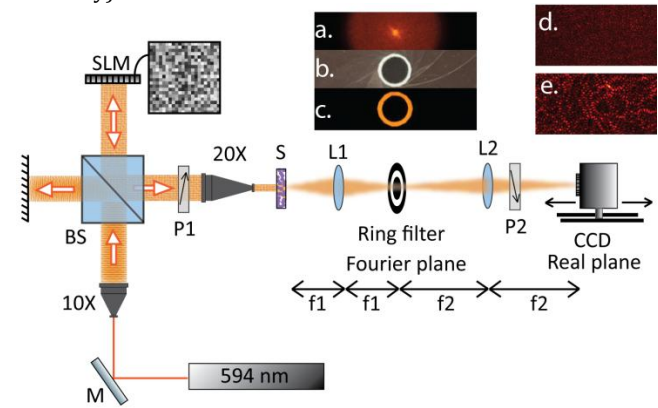

Fig. 1. Experimental setup: the wavefront of a collimated beam is shaped by a spatial light modulator (SLM). Once the beam is de-magnified by a 20X beam reducer, it impinges on the scattering slab S. The output at the back of the curtain (see at the inset d.) is imaged on the pane of the camera by a telescopic system composed by the lens L1 and the lens L2. The lens L1 generates the Fourier Transform (FT) of the back of the sample (see at inset a.) on a plane at its focal distance 25,4mm. When a ring filter (see inset b.) is inserted on this plane we select the portion of the light which the relative spatial frequencies lay

on the ring aperture (see inset c.). Consequently the pattern on the camera changes from the random (see inset d.) to the amorphous distribution (see inset e.).

A second 20x telescope de-magnifies the beam impinging on the scattering sample (S) with a waist of 0.40mm. The scattering sample consists of a dielectric slab of $TiO_2$ with a thickness of $L$=18μm. Lens L1 collects light at the back of the sample and generates the Fourier Transform of the output (Fig. 1a) onto its focal plane (focal length f1=25,4mm). On this plane a ring-shape spatial filter (RSF) is placed in order to select spatial frequencies in a $\delta r$ region (Fig. 1b and 1c). Lens L2 generates the conjugate of the Fourier plane on the plane of the camera. The camera is mounted on a translation stage oriented along the direction of propagation of the light ($z$ axis). Two polarizers (P1 and P2) with reciprocally perpendicular orientation are used in order to eliminate the ballistic contribution through the sample.

The ring-shape filters are fabricated by placing a circular beam stopper at the back of an iris. The beam stopper consists of a circular film of dry mixture of resin with black dye on a transparent cover slip slice. The film has a thickness of 0.50mm and optical characteristics measured in a spectrophotometer (Lambda 950, PerkinElmer, Waltham-Massachusetts, USA) at 600nm: absorption (A=86%), reflection (R=13.6%) and total transmission (T=0.4%). Once the beam stopper is centered on the aperture of the iris an annular aperture is formed with variable width $\delta r$ depending on the iris opening (Fig. 1b). We used three filters with radius $r$=1.45mm and $\delta r$= 0.25mm; 0.80mm and 1.50mm.

As described above, Lens L2 performs a Fourier transform of a ring-shaped momentum space, populated by a random superposition of plane waves and generating a speckle pattern on the plane of the camera. In the absence of the filter (Figure 2.A) a standard speckle pattern (a distribution of dark and bright regions) is generated. However, when the ring filter is applied instead, the patterns appear to have some underlying structure which is more evident if the filter's aperture $\delta r$ is tightened. Figures 2.IIA, B, C & D present the interference patterns generated on the camera plane under different filtering conditions as shown in Figures 2.IA, B, C and D.

We then studied the corresponding power spectrum obtained by Fourier transforming the speckles of Figures 2.IA, B, C & D, as shown in the series of Figures 2.III. In the absence of the filter the complex distribution of speckle corresponds to a uniform distributed power spectrum (Fig. 2.III.A). The introduction of the ring-shape spatial filter in the Fourier domain generates a change in the speckles' distribution and modifies the power spectrum profile as shown in image III.B. of Fig. 2. In practice, decreasing the width of the aperture, $\delta r$, causes the interference pattern to become correlated (see image III.D. of Fig.2) with the typical short-range order which characterizes the amorphous distribution [22].

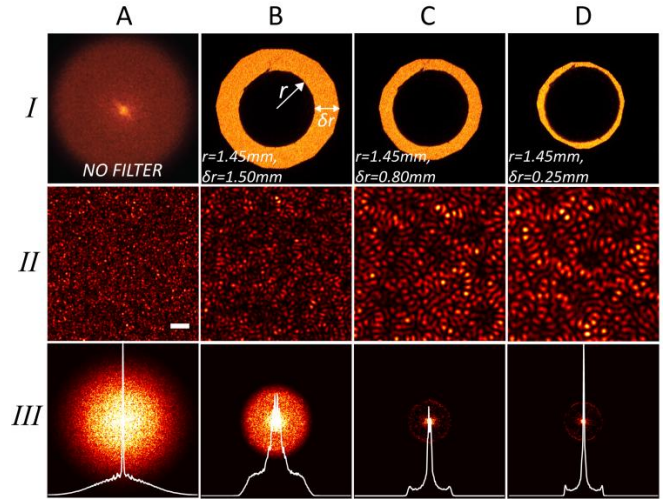

Fig. 2. Images of the different filtering cases used in the study: I.A. no filter; I.B, C & D filters with $\delta r$= 0.25mm; 0.80mm; & 1.50mm, respectively. $r$ is kept fixed at $r$=1.45mm in all cases. II.A, B, C & D the corresponding patterns generated at the back of the filters are compared. The white bar corresponds to 200μm. In III.A, B, C & D we calculate the power spectrum from the corresponding patterns. The superimposed white curves show the logarithm of the intensity profile calculated by radially averaging the power spectrum. Smaller $\delta r$ results to a more defined annular pattern in the power spectrum.

The ring shape in the power spectrum, visible in images III.C. & D. of Fig.2, resemble the characteristic fingerprint of a hyperuniform distribution [23, 24, 25]. Then we want to produce a focused spot exploiting light from these amorphous speckles. This is achieved exploiting an iterative Monte-Carlo algorithm similar to those presented in previous works [27]. A mask composed by 40x40 segments is addressed to the SLM window, we assign to each segment a gray tone on 255 corresponding to a fixed de-phasing of the light reflected by the same segment within a range from 0 up to 5.5π. A preliminary optimization tests a series of 50 random masks picking the one providing the best intensity value at the target position. The mask is taken as input for the second optimization routine, which tests a phase shift of π and -π for a single segment accepting it only if an intensity enhancement on the target is measured, otherwise the previous configuration is restored. Each segment of the mask is tested. The same routine is used for testing a phase shift of π/2 and −π/2 at each segment. At the end of the focusing process we achieve *enhancement h*=300 (ratio between average intensity before and the intensity at the pick after the optimization) in absence of the filter but we obtained much smaller factor when the filter is inserted depending on the size of the aperture [20,21]. As previously demonstrated [4, 10] the great advantage of such approach consists of the ability to generate a spot at user defined positions without mechanically moving the optics. Once the focus is formed at the end of our optimization process we register the position of the camera plane as $z_0$. We collect the extension of the Bessel spot along propagation by exploiting a CCD camera which is driven along the propagation direction (see Fig.1); we then reconstruct the beam profile along the $z$ axis.

In Figures3a), 3b) and 3c) we report images of the Fourier plane generated by L1, the speckle on the camera plane and the focus at the end of the optimization process with its intensity profile (inset white curve), respectively. For this conventional configuration (with NO filter adopted) the shape of the focus obtained via wavefront correction has the same shape of the correlation function of the speckle pattern from which it was initially generated [10]. In the case where a ring-shape filter, with $r$=1.68mm and $\delta r$=0.35mm is present, the final focus obtained at the end of the optimization process presents concentric rings appearing at the region around the central spot as shown in the inset of Fig. 3d and it strongly differs from the case in absence of filter presented in Fig. 3c. We then compare in Fig. 3d the focus intensity profile (dotted black line) to the correlation function of the speckle pattern (dashed green line); again the intensity profile is similar to the correlation function as predicted by Vellekop and co-workers [10]. Furthermore we fit experimental data with a Bessel function profile (solid blue line) and observe that the three curves match consistently each-other; the focus obtained from an amorphous speckle distributions is Bessel shaped. Since Bessel beams are non-diffractive we expect a much larger Depth-of-Focus (*DOF*) with respect to Gaussian beams.

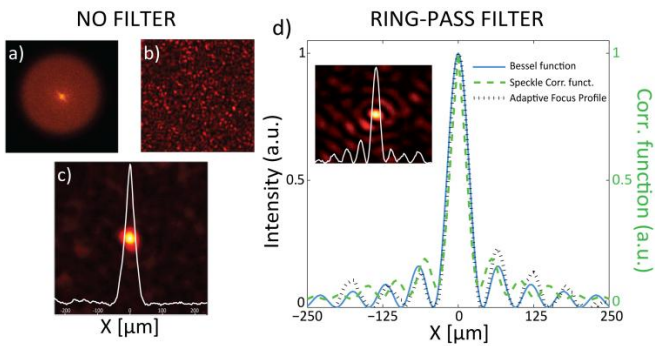

Fig. 3.a) and b) spatial frequencies and speckle pattern produced in the absence of the filter, c) a zoom-in of the focus obtained at the end of the optimization process. d) In the plot is shown the theoretical zero-order Bessel function in blue line, the focus intensity profile in dotted black line and the speckle correlation function of the speckle before the optimization in green dashed line. In the inset it is shown the optimized focus in the presence of the ring filter.

In Fig. 4 we studied the *DOF* (measured as the distance along $z$ for which the focus maintains the same lateral resolution $W_L$ while the Bessel beam intensity is greater than half of the intensity at the peak in the plane $z_0$) for several experimental configurations. The *DOF* is calculated from the *FWHM* of the focus profile along $z$, with the lateral resolution from the *FWHM* along x or y, average on 5 measurements.

In the absence of the ring filter we measured a lateral resolution of $W_L$=23.25μm with a *DOF* of 0.8cm. On the other hand, in the presence of a filter with $\delta r$=0.25mm and $r$=1.12mm, the *DOF* is drastically increased; we obtained a lateral resolution of $W_L$=51.5μm and a *DOF*=29.4cm. In Fig. 4b. each dot represents the intensity at the focus position taken at different planes $z$ starting from $z_0$=0. The effective *DOF* is obtained from the Full Width at Half Maximum of the Gaussian curves obtained by fitting the experimental data. We compared the values of the axial resolutions *DOF* obtained at different $\delta r$: 0.65mm (in green), 0.46mm (in yellow) and 0.25mm (in blue) and with no filter (in red). The curves show that the axial resolution of the focus improves when the ring aperture decreases, the *DOF* ranges from 6.2cm at $\delta r$=0.65mm to 29.4cm at $\delta r$=0.25mm.

As shown in Fig. 4c the beams with lateral waist ($W_L$) are monitored at different ring sizes when propagating along $z$; using very thin ring filters the beam propagates for tens of centimeter without diffracting (solid blue line, $r$=1.12mm) in comparison to a filter of $r$=1.68mm (in red); larger $r$ correspond higher lateral resolution $W_L$ and smaller $\delta r$ correspond to longer beams propagating along z maintaining the same lateral resolution $W_L$ [20,21].

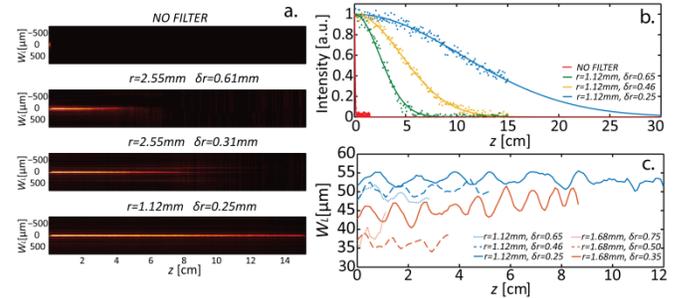

Fig. 4.a) x-z intensity colormap (red = high intensity) of the beam transmitted after the filter. To thinner apertures $\delta r$ corresponds larger the depth of focus. b.) Intensity at the focus as a function $z$ position. resolution is $\delta z$=100μm in the case without filter and $\delta z$=500μm with the filters inserted. Intensity is reported in absence of the filter (red dots), and in presence of filters ($\delta r$=0.65mm (green dots), $\delta r$=0.46mm (yellow dots) and $\delta r$=0.25mm(blue dots)). Continuous line is a fit with a Gaussian curve in order to evaluate the Full Width Half Maximum (*FWHM*) of the intensity profile along z. In c.) the lateral waist ($W_L$) is monitored for different ring sizes when propagating along z.

In summary, we have demonstrated that it is possible to generate non-diffracting Bessel beams through disordered materials by the simultaneous use of adaptive optics, and mode selection by appropriate filtering. The combination of a wavefront shaper, a scattering curtain and a spatial filter results in the Axicon Opaque Lens (AOL), which allows a direct control of both depth of focus and beam waist.

In this study we have used ring-shape spatial filters which perform a momentum selection to the transmitted wavefronts. These wavefronts form an annular speckle distribution in the spatial frequencies domain producing the amorphous pattern, which is non-diffracting and does not diverge [20,21]. As predicted by the model developed by L. Levi and co-workers [20,21] the mean distance between speckles is set by the radius of the ring in momentum space, whereas the symmetry along the propagation, indeed the property to be non-diffractive, is determined by the width of the aperture. Just as standard laser speckles, amorphous speckle patterns can be manipulated by wavefront shaping, in order to enhance their intensity and produce foci at user defined positions. The foci have a non-diffractive Bessel-shape which is equal to the correlation function of the initial speckle.

The reconfigurable non-diffractive beams which we are able to produce through strongly scattering media have the potential to be a an advantage for applied to techniques such as light sheet microscopy or optoacoustic microscopy and mesoscopy, replacing mechanical scanning with a more reliable, stable and fast phase control architecture, as well as the increased depth of field and penetration depth with minimized effective diffusion can dramatically improve *in vivo* imaging.

The authors wish to acknowledge funding from Grants "Skin-DOCTor" and "Neureka!" implemented under the "ARISTEIA"

and "Supporting Postdoctoral Researchers" Actions respectively, of the "OPERATIONAL PROGRAMME EDUCATION AND LIFELONG LEARNING", co-funded by the European Social Fund (ESF) and National Resources and from the EU Marie Curie Initial Training Network "OILTEBIA" PITN-GA--2012-317526.